
\magnification= \magstep1
\tolerance=1600
\parskip=0pt
\baselineskip= 6 true mm

\def\a{\alpha}
\def\b{\beta}
\def\g{\gamma} 
\def\d{\delta} 

\def\h{\eta}
\def\k{\kappa}
\def\l{\lambda} 
\def\m{\mu}
\def\f{\phi} 
\def\n{\nu}
\def\j{\psi} 

\def\s{\sigma}

\def\x{\xi} 

\def\w{\omega}

\def\Lagr{{\cal L}}
\def\pa{\partial}
\def\half{\textstyle{1\over2}}
{\nopagenumbers

\vglue 1truecm
\rightline{THU-93/10}

\vskip 2 truecm
\centerline{{\bf CANONICAL QUANTIZATION OF GRAVITATING POINT PARTICLES}}
\vskip .5 truecm
\centerline{{\bf IN 2+1 DIMENSIONS}}
\vfil
\centerline{G. 't Hooft}
\vskip 1 truecm
\centerline{Institute for Theoretical Physics}
\centerline{University of Utrecht, P.O.Box 80 006}
\centerline{3508 TA Utrecht, the Netherlands}
\vfil
\noindent{\bf Abstract}
\bigskip
A finite number of gravitating point particles in 2+1 dimensions may close
the universe they are in. A formalism previously introduced by the author
using tesselated Cauchy surfaces is applied to define a quantized version
of this model. Special emphasis is put on unitarity and uniqueness of the
evolution operator and on the physical interpretation of the model. As far
as we know this is the first model whose quantum version automatically
discretizes time. But also spacelike distances are discretized in a very
special way.
\eject
}
\pageno=1
\noindent {\bf 1. Introduction}
\medskip
Consider the Lagrangian
$$ \Lagr = {1\over G}\sqrt{-g} R + \sqrt{-g}(-\half g_{\m\n}\pa_\m \f\pa_\n
\f -\half m^2 \f^2) \eqno (1.1)$$

\noindent in 2+1 dimensions. The perturbation expansion in Newton's
constant $G$ is non-renormalizable. Yet one may suspect that a quantum
version of this model exists, because in a very special classical limit the
system is integrable. We have in mind the limit where the $\f$ particles
with mass $m$ become classical gravitating point particles [1]. There are no
gravitons in 2+1 dimensions [2,3]. The particles move in rectilinear orbits
in a locally flat space, and this motion is non-trivial only because the
continuation of these orbits depends on an element of a braid group.

But the small distance behavior of a quantum field theory described by
(1.1) must be very peculiar. Newton's constant $G$ defines a Planck length,
and at this length scale any perturbative approach will break down. Typical
quantum gravitational effects must be essential there.

Now the {\it pure} gravity system, without particles, can be quantized [4].
But adding spinless point particles is essentially equivalent to adding a
scalar field, and this may provide us with an infinite dimensional Fock
space. If second quantization should be inevitable we should prepare for
creation and annihilation of particles, and this was not considered in [4].

In previous papers [5] we expressed doubt whether a rigorous quantum version
exists at all, because of the requirement of a fundamental quantized gravity
theory. But further examination of the classical system provided us with so
much information concerning its fundamental degrees of freedom and its
causal structure that a renewed attack is made possible. We here report on
a beautiful short distance structure, showing features of finiteness that
were previously speculated upon for the 4 dimensional theory.

Our approach for the classical system will be that of Ref. [3]. Let us
briefly summarise the method. We start with a set of Cauchy surfaces $C$ .
A Cauchy surface is a spacelike cross section of space-time. Here it is 2
dimensional. No pair of points in a Cauchy surface is allowed to be
timelike separated, and all other points in space-time must be either in
the future or in the past of at least one point in each Cauchy surface. To
each of our Cauchy surfaces $C$ we attach a time parameter $t_C$ .

Next, we design a tesselation of each Cauchy surface, so that its evolution
to the future (and to the past) can be calculated. Each of these "tiles" is
a polygon. Particles may only sit at the corners of the polygons, so inside
each polygon space-time is flat. The constant time surface $t=t_C$ defines
a preferred Lorentz frame (but not yet its origin) for space-time there.
The polygons are bounded by edges. At an edge, the two Lorentz frames of
the adjacent polygons are related by a Lorentz transformation. Since at an
edge the two time coordinates coincide the Lorentz boost from one polygon
into the next must be directed orthogonally to this edge. The boost parameter
for an edge $L_i$ will be called $2\h_i$. For reasons that will become
clear later we will now choose the signs such that $\h_i>0$ if both
polygons contract; $\h_i<0$ if they expand. The velocity of the edge itself
in both frames is described by half this boost, $\h_i$ . A {\it particle}
at a corner of a polygon will connect two edges that will be glued together
in such a way that the particle produces a conical singularity. In general
the particle moves, so that the Lorentz frames at both sides of such an
edge will also be related by a Lorentz transformation. If $2\b$ is the
deficit angle at a particle, $m$ is its mass, and $\x$ the boost parameter
for its velocity in the local frame, then we have [3]

$$ \eqalign{\tan \b &= \cosh\x\,\tan \half m\,\,, \cr
\tanh\h &= \sin\b\,\tanh\x\ , \cr
\cos \half m &= \cos\b\,\cosh\h\ ,\cr
\sinh\h &= \sin \half m\,\sinh\x\ .\cr}\qquad\qquad\eqalign{(&a)\cr (&b)\cr
(&c)\cr (&d)\cr }\eqno (1.2)$$
(for future calculations it turned out to be convenient to absorb here a
factor $2\pi$ in the definition of the masses $m$ of the particles, as
compared to our earlier expressions in ref [2-5]).

The topological structure of a tesselation will be denoted by a diagram
indicating the edges of the polygons without bothering about actual lengths
or angles. Depending on the global topology of 2-space the diagram should
be seen as living on a topologically non-trivial sheet, which we unfold by
removing a few points. The diagram (after adding the point(s) at infinity)
indicates how the polygons fit together.

The evolution is now indicated diagrammatically.  During short intervals of
time we may simply allow time to evolve equally fast on all polygons, so
that the edges move with their well-defined velocities. But it will be
unavoidable that as time continues something will happen. It could be that
the length of an edge shrinks to zero.  It could also happen, since many
polygons are not convex, that one of the vertices of a polygon hits one of
the other edges, at which point also it becomes illegal to continue the
description in terms of these particular polygons. A transition in terms of
another set of polygons takes place.  It is the succession of many such
transition that we studied. The complete set of all possible transitions in
a diagram is listed in Fig.~1.

\midinsert
\vskip 7.4 truecm \special{berk1.ps}
\narrower Fig.~1. The nine different possible transitions diagrammatically
\endinsert

In most cases a new edge is created, which implies that two polygons that
were truly separated before, now will acquire an edge in common, whereas
other edges may disappear. Since the relative Lorentz transformation
between one polygon an an adjacent one was determined by the succession of
Lorentz transformations at other edges, and since this will not change, one
will always be able to compute both the orientation of the new edge $L_1$
relative to the others and the new Lorentz boost parameter $\h_1$.

In practice we compute these new numbers using "triangle relations".
Consider a vertex between three polygons, $I$, $II$ and $III$, and let
$\a_{1,2,3}$ be the angles between two edges in each polygon, and
$\h_{1,2,3}$ the three Lorentz boosts, labled as shown in Fig.~2. We define

$$ \sin \a_i = s_i\ ,\ \cos \a_i = c_i \ ,\ \sinh 2\h_i = \s_i\ ,\ \cosh
2\h_i =\g_i \ ; \eqno (1.3) $$
\noindent then we have the relations

$$\eqalignno{s_1 :s_2:s_3&=\s_1:\s_2:\s_3\ ,&(1.4)\cr
\g_2s_3+s_1c_2&+c_1s_2\g_3=0\ ,&(1.5)\cr
c_1 =c_2c_3&-\g_1s_2s_3\ ,&(1.6)\cr
\g_1 =\g_2\g_3&+\s_2\s_3c_1\ ,&(1.7)\cr
\cot \a_2 = -\cot \a_1\, \cosh 2\h_3&-\coth 2\h_2\,\sinh2\h_3/\sin \a_1\
,&(1.8)\cr}$$
\noindent and all cyclic permutations.
\midinsert
\vskip 3.8 truecm \special{triangle.ps}
\narrower Fig.~2. The angles $\a_i$ and the boosts $\h_i$ at a vertex
between three polygons

\endinsert

The use of these relations is described in detail in Ref. [3]. Any possible
ambiguity in the parameters of a newly opened edge is removed by requiring
that the edge grows with a positive time derivative and that the complete
set of polygons must form a true Cauchy surface at all times. An edge $L$
grows or shrinks at its two end points $A$ and $B$ :

$$ \dot L = g_A + g_B\ ,\eqno (1.9)$$
At a vertex $A$ the growth $g_A$ of edge $L_1$ is given by
$$g_{A,1} = (v_1\cos\a_3+v_2)/\sin\a_3=(v_1\cos\a_2+v_3)/\sin\a_2\
  ,\eqno(1.10)$$
\noindent where
$$v_i = -\tanh \h_i = -\s_i/(1+\g_i)\ .\eqno (1.11)$$
At a particle $P$ the contribution to the time dependence $\dot L$ is
$$g_P = \tanh \h\,\cot \b\ =\ \tanh \x\,\cos\b\ .\eqno(1.12)$$

\noindent The equations (1.11) and (1.12) show how the edges evolve.

The degrees of freedom of the system are essentially the collection of
lengths $L_i$ of the edges $i$ and the Lorentz boost parameters $\h_i$ at
these edges. The orientations of these edges, and with them the orientation
of the Lorentz boosts there, are then fixed because one can compute the
angles $\a_i$ using first (1.7), after which any ambiguity for the sign of
$s_1$ can be lifted using (1.4) together with the information that at each
vertex at most one of the $s_i$ is allowed to be negative.

There will however be {\it constraints}. Each polygon must close exactly,
which implies that the angles at its $N$ corners must obey

$$ \sum_{i=1}^N (\pi-\a_i) = 2\pi\ ,\eqno (1.13)$$
\noindent (counting the contribution of a particle $P$ as $\a_P=2\pi -
2\b$). Furthermore the vectorial sum of all edges must coincide with the
origin:
$$ \sum_{i=1}^N L_ie^{i\w_i}=0\ ,\eqno (1.14)$$
\noindent where $\w_i$ is the orientation of the edge $L_i$ in the frame of
the polygon, to be computed from the angles $\a_j$ . So each polygon
produces three constraints altogether.

\bigskip

\noindent {\bf 2. Brackets}
\medskip
\nobreak
We have the complete set of degrees of freedom, their equations of motion
(1.9) --- (1.12), and the constraints (1.13) and (1.14) on them (which are
automatically preserved by the equations of motion). Naturally, if we wish
to find a quantum version of this model we have to find a Hamiltonian and
Poisson brackets that generate these equations of motion. The following
construction was discovered by first studying the weak gravity limit, at
which space-time becomes completely flat, and the particles form a Fock
space with known expressions for energies, momenta and Poisson brackets. In
this limit the polygons form diagrams such that particles and clusters of
particles are each connected with lines that are oriented in such a way
that they are all parallel to the total momentum they carry. We read off
from (1.2a) that the deficit angle $2\b$ for a particle coincides precisely
with its energy in this limit (since $m$ is infinitesimal), and from (1.2b)
we see that then $2\h$ precisely corresponds to its momentum.

A complication, of course, is that we are dealing with a cosmology, and
this implies that the total Hamiltonian will have a fixed value. It is
natural now to take as an Hamiltonian the combined deficit angles.  More
precisely, the total energy enclosed inside any closed contour $C$ is the
deficit angle obtained when we parallel transport the local coordinate
frame along this curve.

If this is taken to be the Hamiltonian then we can deduce the canonical
variable conjugated to the length $L_i$ of an edge $i$ by requiring
$$\dot L_i = \{H,L_i\}\ .\eqno (2.1) $$
We know that this variable must be a function of the $h_j$, the boost
parameters of all edges. In the weak gravity limit the variable canonically
conjugated to $L_i$ simply turned out to be $2\h_i$ . In principle one could
have expected a more complicated function of the $\h_j$ in the strong
gravity case. Suppose now that the variable conjugated to $L_i$ is some
function $p_i\{(\h_j)\}$ . Let us then compute the Poisson bracket (2.1).

A particle $P$ contributes to the Hamiltonian $H_P = 2\b$ . Therefore it
contributes to the time derivatives $\dot L_i$ as follows:
$$\d \dot L_i = \{H_P,L_i\}  = \pa H_P/\pa p_i = 2(\pa \b/\pa \h)(\pa \h/\pa
p_i)\ .\eqno (2.2)$$
Eq (1.2c) gives the relation between $\b$, $\h$ and $m$ , from which it
follows that
$$ {\pa \b\over \pa \h}={\sinh\h\over \cosh^2\h}\ {\cos m\over
\sin\b}=\tanh\h\,\cot\b\ .\eqno (2.3)$$
We see that this is exactly the velocity $g_P$ derived earlier, eq.~(1.12).
Hence the contribution of a particle to the time derivative of its cusp
agrees with the Poisson bracket only if the variable $p_i$ canonically
associated to $L_i$ is exactly $2\h_i$ .

Our scheme will only be self consistent if also the contribution of the
vertices as given in eq.~(1.10) agrees with the Poisson bracket (2.1). We
expect that the contribution to the Hamiltonian from a vertex is
$$\eqalign{H_V=H_1+H_2+H_3+2\pi\ ;\quad& H_i=-\a_i\,;\cr
\cos(H_i)={\g_i-\g_j\g_k\over
\s_j\s_k}\,,\quad i,j,k=1,2,3\,.\cr}\eqno (2.4)$$
where we used the triangle relation (1.7).

This will give
$$\eqalign{ \{H_1,L_1\}&=-{\half}\pa \a_1/\pa\h_1={\s_1\over s_1\s_2\s_3}\,;\cr
\{H_2,L_1\}&=-{\half}\pa\a_2/\pa\h_1={\g_3-\g_1\g_2\over
s_1\s_1\s_2\s_3}\,;\cr
\{H_3,L_1\}&=-{\half}\pa\a_3/\pa\h_1={\g_2-\g_1\g_3 \over
s_1\s_1\s_2\s_3}\,;\cr
\{H_V,L_1\}&=\sum_{i=1}^3\{H_i,L_1\}={\g_1+1-\g_2-\g_3 \over
s_2\s_3(1+\g_1)}\,.\cr}\eqno (2.5)$$
Indeed, substituting eq.~(1.7) for $\a_2$ or $\a_3$ in eq.~(1.10) for the
velocity of the edge $L_1$ at the vertex $V$ we find
$$g_{V,1}=-{1\over s_2}\big({c_2\s_1\over 1+\g_1}+{\s_3\over 1+\g_3}\big) =
{\g_1+1-\g_2-\g_3\over s_2\s_3(1+\g_1)}\,.\eqno (2.6)$$

We conclude from eqs.~(2.3), (2.5) and (2.6) that indeed the variables
$2\h_i$ are the canonical conjugates of the $L_i$:
$$ \{2\h_i,L_j\}=\d_{ij}\ .\eqno(2.7)$$
The Hamiltonian is the sum of the deficit angles, as given by (2.4)
for the vertices.
\bigskip
\noindent{\bf 3. Constraints}
\medskip
\nobreak
At every polygon we need to impose the condition that all angles add up to
$2\pi$ as given by the constraint (1.13). The angles of the polygon
contribute to the Hamiltonian. Apparently we have
$$ \sum_{i=1}^N H_i = 2\pi(1-N)\ ={\rm fixed}\ ,\eqno (3.1)$$
where $i$ lables the $N$ corners of the polygon.

The physical interpretation of this constraint is not difficult to see.
Inside each polygon we had been free to choose the Lorentz frame, and in
particular the time coordinate for this local frame. If we allow this
polygon to evolve all by itself it is governed by this part of the
Hamiltonian. The constraint (3.1) tells us that this is an invariance of
the state of the system.

The complex constraint (1.14) must correspond to invariance with respect to
Lorentz transformations of the frame inside the polygon. The effect a
Lorentz transformation inside one polygon has on the surrounding $L_i$ is
rather complicated, so we have not checked explicitly whether the
change generated by this constraint indeed matches this.

Besides these there are however more subtle constraints, in the form of
{\it in}equalities:
$$L_i \geq 0\ ,\eqno(3.2)$$
(obviously the lengths of the edges must be greater than or equal to zero).
{}From (1.7) one can also deduce that the $\h_i$ must satisfy a {\it triangle
inequality}:

$$ |\h_i| \leq |\h_j|+|\h_k|\,,\quad i,j,k=1,2,3\ .\eqno (3.3)$$

We here anticipate that another set of canonical variables will be useful:
$$ x_i=L_i\, {\rm sgn}(\h_i)\,,\quad p_i=|2\h_i|\ ,\eqno(3.4)$$
which, at least classically, also obey the Poisson brackets
$$\{p_i,x_j\} = \d_{ij}\ .\eqno (3.5)$$
Note that the question whether the commutator analogues of (2.7) and (3.5)
are also equivalent will be not so obvious.

The advantage of these variables will be that there is now no further
constraint on (the sign of) $x_i\,$, whereas $p_i$ are now limited to be
positive. In addition the $p_i$ satisfy the ordinary triangle inequality,
(3.3) without the absolute value signs.

\midinsert
\vskip 3.7 truecm \special{overlap.ps}
\narrower Fig.~3. Acceptable self-overlapping polygon
\endinsert

There still is another inequality which is more difficult to write down
explicitly. This is the requirement that all pogygons must be {\it true
polygons}. The point here is that the angles of a polygon need not be
convex. In particular a particle with mass less than $\pi$ will give a cusp
corresponding to an angle larger than $\pi$ for the polygon. Consequently
we must be careful that the polygon does not intersect with itself,
otherwise our surface ceases to be a genuine Cauchy surface. In the
classical case, if a polygon tends to disect itself it will undergo a
transition of the type depicted in Fig.~1c or 1d.

Occasionally however we will have a polygon that does overlap with itself;
yet it may still be an acceptable polygon if it is still a boundary of a
two-dimensional space, see Fig.~3. It only overlaps with itself if we insist
on choosing a rectangular coordinate grid within.
\bigskip
\noindent{\bf 4. Quantization}
\medskip
\nobreak
In this section the commutators will be replaced by $-\hbar i$ times the
Poisson brackets. We now claim that the variables $x_i$ and $p_i$ are to be
preferred rather than the $L_i$ and $2\h_i$ as canonical variables. This is
because at $L_i =0$ a transition of the form of Fig.~1a, b, e, \dots, j has
to take place, which corresponds to adding a boundary condition at $L_i=0$.
When a new edge opens up it is not a priori clear how to resolve the sign
ambiguity in the determination of the new $\h$ variable, and in association
with that the signs of the trigonometric functions sine and cosine of the new
angles. If we replace the $L_i$ by quantities $x_i$ we can use the signs of
$x_i$ to give these lines an orientation inwards or outwards of the vertex.
In particular at the transitions of Fig.~1e and f the newly resulting wave
function will be continuous in $x$ (note that in these two figures the signs
of $\h$ on the horizontal line segments flip).  The advantage of keeping $p_i$
positive will become apparent in Section 5.

The transitions as pictured in Fig.~1 should be seen as boundary conditions
on the wave function. But we can also view them as providing identities for
the wave function on different diagrams. If the wave function is known on
any particular diagram we can derive it on any other diagram by using these
identities. For instance if in Fig.~1a the edge shown at the left has a
length $L>0$ then the diagram at the right may be seen as an analytic
continuation of it, such that the new length parameter $L <0$. Figs.~1c, d
and 1g---h show how polygons can be added to or removed from a diagram.

Fig.~1 was originally intended to list the transitions for a classical
theory, not directly to formulate the quantum system. For that it will
probably be more convenient to reformulate the rules slightly. Technically,
the transitions c and d can be obtained most easily by first aplitting a
polygon in two, using a new edge with $\h=0$. Since then also $p=0$ this
simply means that the wave function does not depend on the new $x$ variable
at all, and so this is a dummy variable at this stage. But then transitions
of the type a and b are performed, and after that we obtain non-vanishing
$\h$ and hence non-trivial dependence on the new $x$ variables.

Also, adding or removing a polygon by transitions of the form of Fig.~1g or
h is straightforward. Because of the Hamiltonian and Lorentz constraints on
the extra polygon the wave function depends neither on the Lorentz
orientation nor on the internal time variable of that polygon.

Let us stress once again that it is the transitions that cause our system
to be highly non-trivial. The classical system already has shown that
infinite successions of such transitions often occur (usually resulting in
ever increasing values for $p_i$ , in which case the $L_i$ equally rapidly
decrease. So the consequences of the constraints induced by the boundary
conditions of Fig.~1 are severe.

Ultimately, since we are performing cosmology rather than local quantum
mechanics it is not so much the Schr\"odinger equation but rather a
Wheeler-DeWitt type equation to which the entire wave function will obey.
\bigskip
\noindent{\bf 5. Discreteness}
\medskip
\nobreak
The most striking consequences of the quantum structure of this model for
some reason has never been observed or stressed by other authors. It is the
discreteness of the relevant variables. First let us concentrate on the
time variable.

The Hamiltonian is the total deficit angle. For a closed $S_2$ universe
this is $4\pi$. Locally, the contributions to the Hamiltonian govern how
parts of the universe evolve with respect to a time variable fixed at
"distant polygons", or "distant observers". It seems that the very physical
nature of our approach allows us to see this more clearly than otherwise.
What we see is that the local Hamiltonians are also angular. Of course
these angles are defined only {\it modulo} $2\pi$ , and so our Hamiltonians
are also only defined {\it modulo} $2\pi$ . Indeed, all expressions we have
for the Hamiltonians in terms of the $p_i$ give us only $\cos H_i$ (eq.~
1.7), and to some extent also $\sin H_i$ (eq.~1.4). This means that what we
really have is only direct expressions for $e^{\pm iH}$ in terms of
single-valued functions of $p_i$.

But this we consider as highly interesting. Apparently the evolution of the
system is only well determined for integer time steps. Fractional time
steps are ill-defined and skipped. Clearly time is quantized in our model.
In fact, this quantization of time was seen earlier when the relation
was established between angular momentum on the one hand and a time shift
along a contour around a set of particles on the other. Since angular
momentum is quantized, time shifts are quantized also [5].

Time quantization is also essential for a discussion of uniqueness and
unitarity of the system. We want the evolution to be described
unambiguously. One can only hope to obtain such an unambiguous law for
time steps that are integer. Yet even there things are not quite this
simple. Eq. (1.7) does give us the cosine of
the contribution to the Hamiltonian, but not the sine. This means that,
from that equation alone, we obtain the operator
$$ e^{-iH} + e^{iH}\,,\eqno (5.1)$$
a combination of a step to the future and a step to the past. What is
needed is an unambiguous expression for $e^{-iH}$ alone.

Fortunately we also have (1.4). In the classical system this expression is
sufficient to determine all angles uniquely at every transition. Given the
$\h_i$ there is in principle one overall sign ambiguity for the sines of
the angles at each vertex. Of course what this means physically is that our
quantum system evolves under laws that ere symmetric under time reversal.
All that is needed is that for one diagram all angles must be given. We can
then use eq.~(1.4) to lift the sine ambiguity for all transitions to all
other diagrams.

To establish whether our system will be copletely unitary is still
difficult however. If we take a complete cosmology we know that large
series of transitions can relate one diagram to the same diagram at many
different values of the parameters. So we cannot allow for all possible
states $\j({\bf x})$ because most of them will violate the boundary
conditions of Fig.~1, or, they will violate the Wheeler-DeWitt equation.
It is at a local level that our equations seem to be sufficiently stringent
to determine $e^{-iH}$ completely.

Since time is discrete one may now ask to what extent space is discrete
also. It would be natural, perhaps, to suspect that space-time forms a
complete lattice. But the situation is considerably more complicated. In
principle the $x_i$ can take any (integer or fractional) value. The thing
to observe however is that the only $p$ dependence comes from the
hyperbolic sines and cosines of $p_i$, or, we just see the operators $e^{\pm
p_i}$ occurring in our expressions both for the cosines and for the sines of
$H_i$ . But these are the operators for shifts over exactly a distance
$i$~, which is a unit step in the imaginary direction. The question now is
how to exploit this fact.

It is now of importance to use the constraint from (3.4) that $p_i$ are
non-negative. This means that the wave functions will be analytic in the
entire upper half plane, Im$(x_j) \geq 0$ .

We have (we omit the indices $j$)
$$ \j(z)={1\over 2\pi i}\int_{-\infty}^\infty {\j(x){\rm d}x\over x-z}\ ,\
{\rm if\ Im}\,z >0\,.\eqno(5.2)$$

Let us now define the amplitudes
$$ \f_n=\j(in)={1\over 2\pi i}\int_{-\infty}^{\infty}{\j(x){\rm d}x\over
x-in}\,\quad (n>0)\,;\quad \f_0=\j(0)\ .\eqno(5.3)$$
Then the operators $e^{\pm p}$ act on these in a very simple way:
$$ e^{\pm p}\f_n = \f_{n\mp1}\ .\eqno(5.4)$$

We are now in a position to interpret the eqs.~(1.4---8) as difference
equations on the wave functions $\f_{\{n_j\}}(\{t_F\})$ . There is a time
variable $t_F$ at each polygon $F$ , which we are allowed to vary separately.
In a given diagram each edge $j$ has a (non-negative) variable $n_j$ . We
lable the polygons and the edges in a cyclically symmetric way as in Fig.~2.
There is now an algebra of operators. At each corner $i$ there is an
operator $T_i=e^{-iH_i}$ , defined such that
$$ T_F=\prod_{i\in F} T_i = e^{-iH_F}\eqno(5.5)$$
is the time displacement operator at polygon $F$:
$$ T_F \f_{\{n_j\}}(t_1,\dots,t_F,\dots )=\f_{\{n_j\}} (t_1,\dots,t_F+1,\dots
)\,.\eqno
(5.6)$$
And at each edge $j$ there is an operator $U_j=e^{p_j}$ defined such that
$$ U_j \f_{ n_1,\dots,n_j,\dots}(t_1,\dots )= \f_{ n_1,\dots,n_j-1\dots}
(t_1,\dots)\ .\eqno(5.7)$$
The entities $s_i,c_i$,$\s_j,\g_j$ in eqs (1.4---8) are now
$$\eqalign{s_i=-\half i(T_i-T_i^{-1})\,,\qquad&c_i=\half(T_i+T_i^{-1})\,,\cr
\s_j=\half(U_j-U_j^{-1})\,,\qquad&\g_j=\half(U_j+U_j^{-1})\,.\cr}
\eqno(5.8)$$

Because of the observed analyticity in the upper half plane we have as a
boundary condition
$$ \f_{\dots,n_j,\dots}\rightarrow 0\quad {\rm as }\quad n_j\rightarrow+
\infty\ .\eqno(5.9)$$
Now most of our equations will be of second degree (that is, involving at
least two steps) in the $n_j$ direction, so that an other boundary
condition may be needed. This is the value of the wave function $\j$ at the
origin. Here an edge vanishes, and hence the wave function must coincide
with other wave functions for different diagrams, to be obtained via
transitions as given in Fig.~1. This is the reason why in eq.~(5.3) we
considered only the set of wave functions that is connected to the origin
by integer vertical steps. These are probably more essential than the ones
we would have obtained had we started at an arbitrary other point in the
complex plane.

Finally we note that the equations at an edge to which a particle is
connected, eqs (1.2) must be treated exactly as eqs (1.4---8). The mass $m$
is an arbitrary free but fixed parameter here. Only one problem has not yet
been addressed. This is the fact that, since the distances are now
discrete, the distance between two particles can become exactly equal to
zero. This was never a concern in the classical case, because such an event
would occur with probability zero. Now it is a finite possibility.
Presumably there will be room here to enter non-trivial non-gravitational
interactions among the particles themselves. this we have not yet worked
out.
\bigskip
\vbox{\noindent{\bf 6. Discussion and Conclusions}
 \medskip
Our procedure with tesselated Cauchy surfaces turned out to be strikingly
suitable for a desciption of not only classical but also quantized
particles gravitating in 2+1 dimensions. The lengths $L_j$ of the edges of
the polygons and the Lorentz boosts $2\h_j$ across these edges turned out
to be each other's canonically associated degrees of freedom, and the
Poisson brackets, eq.~(2.7), are quite suitable for setting up a
quanization procedure.}

But replacing Poisson brackets with commutators must be done with care.
Often, if a Hamiltonian is not quadratic in the momenta, a theory may turn
out to become non-local, non-unitary or lacking a stable vacuum state.  In
this model we faced all these dangers. Now there seems to be a general rule
that if a model is classically integrable it will have an integrable
quantum version as well.  It seems that this rule works to our
advantage here.

The effect of the quantization procedure is remarkable. Because the
Hamiltonian is an angular variable the time coordinate comes out
automatically being discrete. Only over integer time intervals the wave
function evolves unambiguously. Because of Lorentz invariance one could
have expected therefore that also the spacelike dimensions should be
discrete. Instead, it is the {\it imaginary} parts of spacelike distances
which will be quantized in integers. We found that the independent
variables $t_F$ and $n_j$ in the wave function $\f_{\{n_j\}}(\{t_F\}) $
can all be
restricted to integer values, because the wave equations in terms of these
variables turned into difference equations.

We have here an unusual analogue of the so-called Wick rotation in quantum
field theory: instead of replacing time by an imaginary quantity we have
kept time real but replaced the spacelike coordinates by imaginary numbers.
Of course in both cases space-time obtains a (locally) Euclidean signature.

Whether the variables $x_j$ can be completely replaced by the $n_j$ remains
to be seen. This replacement appears to become inefficient at large
distances. A study of the possible classes of analytic wave functions reveils
that there exist indeed functions that are zero on $z=+in$, $n \geq 0$ and
finite when $z$ is real. These are necessarily divergent on the lower half
of the complex plane. Therefore the functions $\f$ are not completely
representative for all states. One cannot build a complete basis out of them.

A complete formulation of "quantum cosmology" in 2+1 dimensions has not yet
been given. What we would like to see is an $S$-matrix construction: given
some asymptotic states $|\j\rangle_{\rm in}$ at time $t\rightarrow -\infty$
(if the universe started being infinitely large) or $t=t_0$ (if the
universe started with a big bang at $t=t_0$), and $|\j\rangle_{\rm out}$ at
time $t\rightarrow +\infty$ (for an expanding universe) or $t=t_1$ (for a
universe with a "big crunch" at $t=t_1$), we would like to be able to
compute the "scattering matrix"
$${ }_{\rm out}\langle\j|\j\rangle_{\rm in}\ .\eqno (6.1)$$
A problem here is to give a semiclassical description of the asymptotic
states. This seems to be all right if the universe became infinitely large
there, but in the crunching case this is very problematic. In an earlier
paper we expressed the suspicion that the crunching states become
semiclassical also. This however was based upon the hope that the momenta
associated with the lengths $L_j$ were something like the hyperbolic sines
or cosines of the $\h$ variables, which was not so crazy because the classical
expressions for the momenta do contain the Lorentz  $\g$ parameters. Now
we know that this is different in the strong gravitational
case, and the asymptotic states will keep their fundamental
quantum mechanical nature.

We actually found from the triangle equations (1.4---10) that
$$\sum_{i=1}^3 g_i\h_i\eqno (6.2)$$
is strictly bounded, even as $\h\rightarrow\infty$.
One can conclude from that that during a classical crunch the quantity
$$\sum_i \h_iL_i\eqno (6.3)$$
approaches a {\it finite} limit. Since this quantity counts the number of
wave nodes one can deduce that the asymptotic crunching state cannot be
described semiclassically. Therefore it will be very difficult to even
define the matrix (6.1). We do not know how to characterize complete
sets $|\j\rangle_{\rm in,out}$ without overcounting.

{}From the fact that (6.2)
approaches a finite limit it also follws that the critical coefficient $\k$
mentioned in Ref. [7] is equal to one.

An alternative approach to a physical interpretation of the quantum theory
could be to concentrate on the definition of an $S$-matrix for scattering
in an open universe, as was done in Ref. [5]. In that case the asymptotic
states are always expanding, but the total energy must be constrained to be
less than $2\pi$. Such a constraint would make it harder however to formulate
the usual conditions of unitarity and causality for the $S$-matrix.

Discreteness of time has the consequence that energy is only defined {\it
modulo} $2\pi$. We could call this unit the Planck energy. In other theories
with discretized time this is a problem, because then there may not be a
well-defined stable vacuum state [6]. Here we are not so much concerned
with that. The {\it total} energy of the universe is only $4\pi$, or two
Planck units.
So if we take a small section of this universe then the energy quantum is
much too large to cause us concern.

Can one add non-gravitational interactions to the model? What about a
$\l\f^4$ term in the Lagrangian? We must observe firstly that, although we
seem to have a completely quantized model here, we have not yet seen creation
or annihilation of particles. We know that creation and annihilation do occur
in a non-gravitaional theory with a $\l\f^4$ interaction. One should
suspect this still to happen if one then adds gravitation. Secondly, so far
we ignored the  states where two particles coincide. Just because of the
discreteness of the distances $L_j$ this may be a serious omission that has
yet to be addressed. Notice furthermore that the
relevant equation linking energy and momentum, eq.~(1.2c), only contains
$\cos m$~, not $\sin m$~. So one may easily generate a sign difficulty for
the particle mass $m$, comparable to difficulties that led to the necessity
of second quantization in non-gravitational field theories. Since the
transitions of Fig.~1b, c and e leave no
ambiguity for the deficit angle corresponding to the Hamiltonian we do not
expect particle creation or annihilation to occur in pure gravity. But as
soon as other interactions are incluided we probably will have to deal with
a complete Fock space.

\bigskip

\centerline{\bf References}
\nobreak
\medskip
\item {1.} A. Staruszkiewicz, {\it Acta Phys. Polon.} {\bf 24} (1963) 734;
\item {\ } J.R. Gott, and M. Alpert, {\it Gen. Rel. Grav.} {\bf 16} (1984)
243;
\item {\ } S. Giddings,  J. Abbot and  K. Kuchar, {\it Gen. Rel. Grav.}
{\bf 16} (1984) 751.
\item {2.} S. Deser, R. Jackiw and G. 't Hooft, {\it Ann. Phys.} {\bf 152}
(1984) 220.
\item {3.} G. 't Hooft, {\it Class. Quantum grav.} {\bf 9} (1992) 1335,
{\it ibid.}, to be publ. (1993).
\item {4.} E. Witten, {\it Nucl. Phys.} {\bf B311} (1988) 46;
\item {\ } S. Carlip, {\it Nucl. Phys.} {\bf B324} (1989) 106; and in:
"Physics,  Geometry  and Topology", NATO ASI series B, Physics, Vol.
{\bf 238}, H.C.  Lee  ed.,  Plenum 1990, p. 541.
\item {5.} G. 't Hooft, {\it Commun. Math. Phys.} {\bf 117} (1988) 685;
\item {\ } S. Deser and R. Jackiw, {\it Comm. Math. Phys.} {\bf 118}
(1988) 495.
\item {6.} G. 't Hooft, K. Isler and S. Kaltzin, {\it Nucl. Phys.}
{\bf B 386} (1992) 495
\item {7.} G. 't Hooft, {\it Nuclear Phys.} {\bf B 30} (1993) 200
\end